\documentclass[preprintnumbers,floatfix,twocolumn,aps,prl,unsortedaddress,superscriptaddress,longbibliography]{revtex4-2}

\usepackage[para]{threeparttable} % for table footnotes
\usepackage{amsmath}
\usepackage{graphicx} 
\usepackage{booktabs}
\usepackage{subcaption}
\usepackage{url}
\usepackage{hyperref}
\usepackage{miller}
\usepackage{bm}
\usepackage{enumitem}
\usepackage{color}
\usepackage{siunitx}

\usepackage[utf8]{inputenc}

\begin{document}

\title{Four regimes of primary radiation damage in tungsten}
%\title{Four energy regimes of primary radiation damage in tungsten}
%\title{Four energy regimes of primary radiation damage production in tungsten}
%\shorttitle{Test}

\author{J. Byggmästar}
\thanks{Corresponding author}
\email{jesper.byggmastar@helsinki.fi}
\affiliation{Department of Physics, P.O. Box 43, FI-00014 University of Helsinki, Finland}
\author{V-M. Yli-Suutala}
\affiliation{Faculty of Science and Engineering, Åbo Akademi University, Vattenborgsvägen 3, FI-20500 Åbo, Finland}
\affiliation{Department of Physics, P.O. Box 43, FI-00014 University of Helsinki, Finland}
\author{A. Fellman}
\affiliation{Department of Physics, P.O. Box 43, FI-00014 University of Helsinki, Finland}
\author{J. Åström}
\affiliation{CSC -- IT Center for Science Ltd., P.O. Box 405, 02101 Espoo, Finland}
\affiliation{Department of Physics, P.O. Box 43, FI-00014 University of Helsinki, Finland}
\author{J. Westerholm}
\affiliation{Faculty of Science and Engineering, Åbo Akademi University, Vattenborgsvägen 3, FI-20500 Åbo, Finland}
\author{F. Granberg}
\affiliation{Department of Physics, P.O. Box 43, FI-00014 University of Helsinki, Finland}

\date{\today}

% shorten!?
\begin{abstract}
{We observe for the first time in silico the transition to a linear regime in the primary damage production in tungsten. As the critical plasma-facing material in fusion reactors, radiation damage in tungsten has been studied extensively in experiments and simulations. Irradiation experiments routinely produce recoils in the MeV range while full atomistic modelling has been limited to a few hundred keV. Here we bridge these scales with extremely large-scale and accurate machine-learning-driven molecular dynamics simulations with recoil energies up to 2 MeV in systems up to one billion atoms. We reveal four regimes of primary damage as a function of damage energy, with a transition to a high-energy regime that deviates from all previous models. Curiously, the start of the high-energy regime coincides with the highest possible recoil energy to tungsten atoms from fusion-emitted neutrons (300 keV).}
\end{abstract}

\maketitle

Formation of primary radiation damage produced by energetic particles is the fundamental mechanism to understand when probing the irradiation resistance of a material. The physics of primary damage formation is quite well understood, largely from detailed analysis of molecular dynamics (MD) simulations~\cite{delarubia_role_1987, stoller_111_2012, nordlund_primary_2018}. Damage by irradiation is possible when the energy transferred to a lattice atom exceeds the material-specific minimum threshold displacement energy. Higher recoil energies cause more atom displacements, producing a cascade of atom collisions and, in the simplest damage models~\cite{kinchin_displacement_1955, norgett_proposed_1975}, a defect production that increases linearly with damage energy (the total ion kinetic energy minus losses to electronic interactions). However, collision cascades in metals evolve into heat spikes of locally dense and hot liquid-like regions, facilitating recombination of the initially displaced atoms during recrystallisation~\cite{delarubia_role_1987}. The current primary damage models account for this with an athermal recombination function to produce a sublinear defect production trend~\cite{nordlund_improving_2018}. At higher energies, cascades split into subcascades with increasing probability and the defect production is assumed to transition from a sublinear to a final linear regime. In materials with a high subcascade splitting energy the final linear regime is, however, difficult to reach by MD simulations. Additionally, some data suggest that in heavy metals the defect production might not simply transition from a sublinear to a final linear trend but include an intermediate superlinear regime that is still poorly understood~\cite{setyawan_cascade_2015,nordlund_improving_2018,fu_molecular_2019,liu_large-scale_2023}. In this Letter we explore, quantify, and explain all these regimes for the first time.

The primary damage in tungsten is of particular importance as tungsten is the leading choice for the plasma-facing wall material of tokamak fusion reactors. Since the emergence of W as the main fusion reactor material, the radiation tolerance has been intensively studied experimentally and computationally~\cite{yi_-situ_2016-1,yi_characterisation_2015,el-atwani_loop_2018,zhang_defect_2016,wielunska_dislocation_2022,wang_dynamic_2023,sand_high-energy_2013,sand_cascade_2017,mason_direct_2018,setyawan_displacement_2015,granberg_molecular_2021}. The 14.1 MeV neutrons emitted from the fusion reactions will produce recoil energies up to 300 keV in the W lattice. MD simulations have been crucial in quantifying and in detail observing the defects of the primary damage. However, the chaotic nature of collision cascades requires large simulations. The largest full MD simulations have now reached 300 keV~\cite{fu_molecular_2019,liu_displacement_2023}, covering the entire fusion-relevant range, although what occurs at even higher energies remains unexplored.

Due to the lack of fusion-like neutron sources, irradiation experiments often use heavy-ion irradiation in the MeV range to emulate the damage produced in fusion reactors and to accumulate high doses quickly. Such heavy ions can produce recoil energies far higher than the 300 keV maximum from fusion-emitted neutrons. For example, Yi et al.~\cite{yi_characterisation_2015} carried out extensive irradiation studies with 2 MeV self-ion irradiations, Wielunska et al.~\cite{wielunska_dislocation_2022} used 20.3 MeV self-ions and 7.5 MeV Si ions (translating to W recoils up to 3.45 MeV), El-Atwani et al.~\cite{el-atwani_loop_2018} used 3 MeV Cu ions (up to 2.3 MeV W recoils), and Wang et al.~\cite{wang_dynamic_2023} used 6 MeV Cu ions (up to 4.6 MeV W recoils). Thus there is a wide gap between the experimentally and the so far computationally reached energy range. Crucially, it has been observed that the defect production by collision cascades in W deviates upwards from the expected linear trend predicted by damage models at recoil energies above 50 keV~\cite{setyawan_cascade_2015,nordlund_improving_2018,fu_molecular_2019,liu_large-scale_2023}. The mechanism behind this superlinear trend was suggested to be related to the proximity of subcascades~\cite{setyawan_cascade_2015}. The authors argued that when cascade splitting into subcascades occurs in heavy metals, the subcascades are often so close to each other that they overlap and result in large defect clusters and hence in more defects than expected~\cite{setyawan_cascade_2015}. The same authors also speculate that at even higher energies, the subcascades should become fully separated and the damage production should return to a linear trend. This transition has so far not been observed and the final high-energy linear trend not quantified. The critical question is at which energy this transition occurs relative to the fusion-neutron recoil spectrum (up to 300 keV) and typical heavy-ion irradiation experiments (MeV range), since it dictates how analogous heavy-ion radiation damage is to neutron-induced damage.

\begin{figure*}
    \centering
    \includegraphics[width=0.9\linewidth]{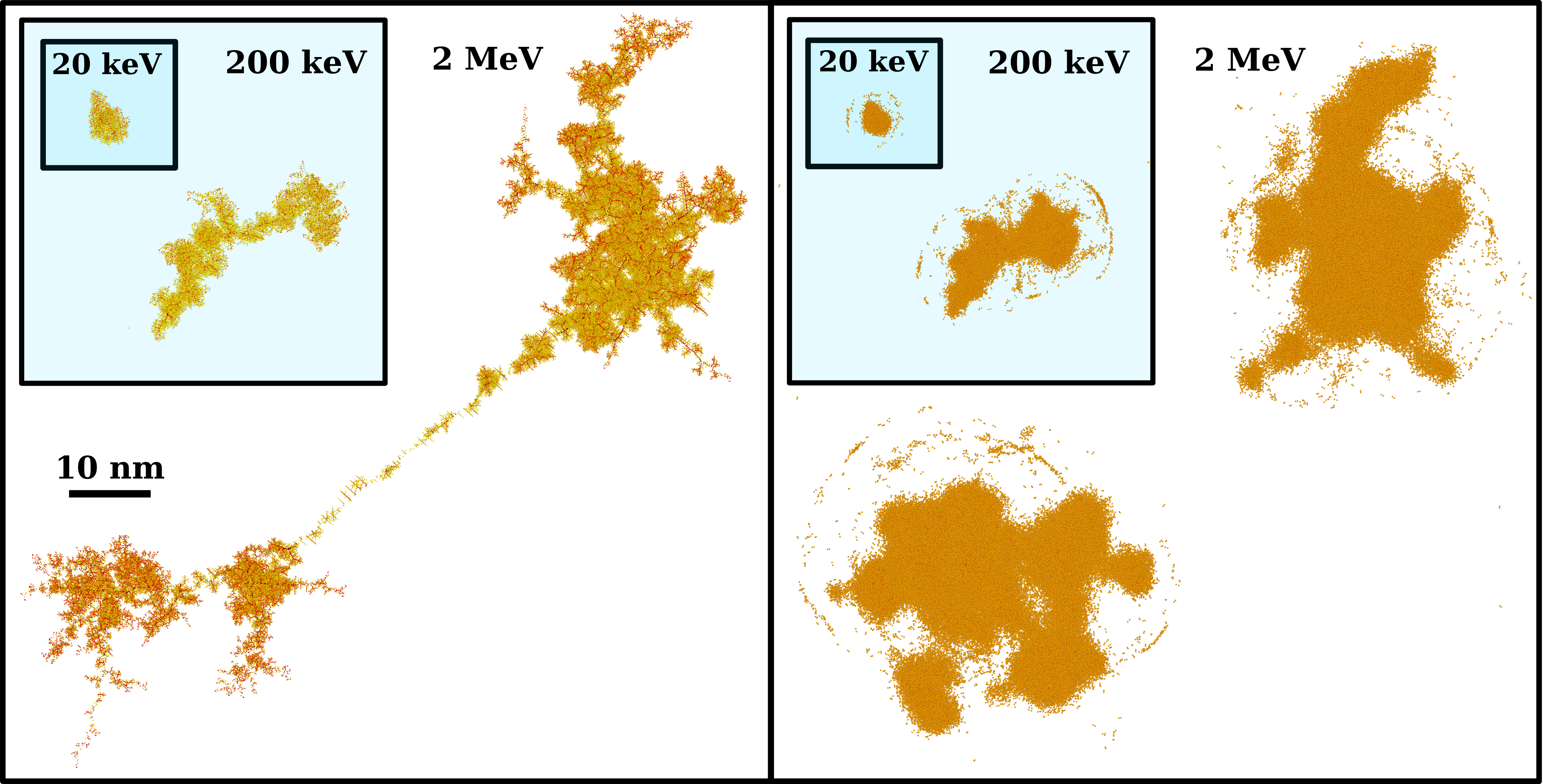}
    \caption{Typical collision cascades across three orders of magnitude in recoil energies. Atoms with kinetic energies above 1 eV are shown and atoms are colored according to their kinetic energy (red: high, yellow: low). On the left, the cascades are visualised as all high-energy atoms up to times just before the heat spike (150, 170, and 190 fs for 20 keV, 200 keV, and 2 MeV, respectively) to show typical structures and sizes. On the right, the same cascades are shown as all high-energy atoms at a time during the heat spike and the emission of pressure pulses (1.0, 1.7, 3.3 ps, respectively), revealing three different types of cascades; single cascade, overlapping subcascades, and fully separated subcascades.}
    \label{fig:casc}
\end{figure*}

\begin{table}
    \centering
    \caption{Primary knock-on energies ($E_\mathrm{PKA}$), the number of atoms in the simulated lattice ($N_\mathrm{atoms}$), the length of the cubic simulation system ($L$), and the number of independent cascade simulations with randomly directed recoils ($N_\mathrm{sim}$). The cascades are simulated as one recoil per simulation unless otherwise noted.}
    \begin{threeparttable}
    \begin{tabular}{lrrr}
     \toprule
     $E_\mathrm{PKA}$ (keV) & $N_\mathrm{atoms}$ & $L$ (nm) & $N_\mathrm{sim}$ \\
     \midrule
     0.04--0.25 & \num{54e6} & 95.6 & 1000\tnote{a} \\
     0.3--1.0 & \num{16e6} & 63.7 & 500\tnote{b} \\
     2, 5 & \num{0.128e6} & 12.74 & 100 \\
     10, 20 & \num{1.024e6} & 25.5 & 100 \\
     50 & \num{8.192e6} & 51.0 & 100 \\
     100 & \num{16e6} & 63.7 & 100 \\
     200 & \num{54e6} & 95.6 & 50 \\
     300, 500 & \num{128e6} & 127 & 50 \\
     1000, 2000 & \num{1.024e9} & 255 & 20 \\
     \bottomrule
    \end{tabular}
  \begin{tablenotes}
   \item[a] A grid of 1000 recoils in one simulation.
   \item[b] Grids of 125 recoils in four independent simulations.
  \end{tablenotes}
 \end{threeparttable}
    \label{tab:sim}
\end{table}

Here we resolve this issue using extremely large-scale and accurate MD simulations with a machine-learned interatomic potential~\cite{byggmastar_simple_2022} to reach unprecedented damage energies with full atomistic resolution. We achieve this by implementing the computationally efficient tabulated Gaussian approximation potential (tabGAP)~\cite{byggmastar_simple_2022} to run on GPU hardware through the Kokkos library~\cite{9485033} and \textsc{lammps}~\cite{thompson_lammps_2022}. We then perform MD simulations with primary knock-on atom (PKA) energies across six orders of magnitude, from 40 eV to 2 MeV, to obtain the primary damage production trend for all practically relevant energies. Containing the collision cascades within the simulation cell required W lattices up to one billion atoms for the highest PKA energies (1 and 2 MeV). The systems are initially relaxed to 300 K and zero pressure. In the cascade simulations, we use an adaptive time step, a total simulation time of 50 ps, and the $NVE$ ensemble for all atoms except for those in an 8 Å thick border region for which the Nosé-Hoover thermostat (300 K) is applied. The border thermostat dissipates heat and dampens pressure pulses from the cascade to avoid unphysical cascade self-interaction. For high energies, the energy loss to electronic excitations becomes significant. We consider electronic stopping as a friction force for atoms with kinetic energies above 10 eV using data from \textsc{srim}~\cite{ziegler_srim_2010}. The PKA energies, system sizes, and number of cascade simulations are summarised in Tab.~\ref{tab:sim}. The directions of the independent recoils are sampled randomly. The simulations are carried out on the GPU nodes of the LUMI supercomputer. The damage is analysed using Wigner-Seitz analysis for finding vacancies and self-interstitial atoms (SIAs), the dislocation extraction algorithm (DXA) for finding dislocations, and cluster analysis to obtain cluster size statistics, all implemented in and visualised using \textsc{ovito}~\cite{stukowski_visualization_2010,stukowski_automated_2012}.

\begin{figure*}
    \centering
    \includegraphics[width=0.9\linewidth]{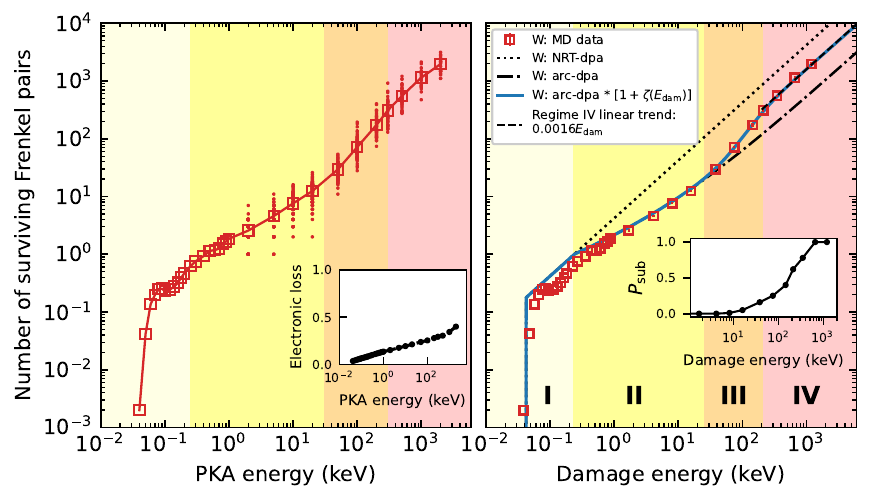}
    \caption{Number of Frenkel pairs formed as a function of PKA energy (left) and damage energy (right). The damage energy is the PKA energy minus the energy loss to electronic stopping, the fractions of which is shown in the inset figure. The highest PKA energy 2 MeV corresponds to, on average, 1.2 MeV damage energy. The right inset figure shows the probability of subcascade splitting obtained from visual analysis of the cascade simulations. The damage production is grouped into four regimes, I: near-threshold energies with on average less than one defect produced, II: single cascades following the arc-dpa model, III: transition region with superlinear damage trend, and IV: the final linear regime observed for the first time in W here. The NRT-dpa and arc-dpa models~\cite{nordlund_improving_2018} with the low-energy correction~\cite{yang_full_2021} are plotted for comparison as well as the final linear trend and the new model that follows the trend across all four regimes.}
    \label{fig:nvacs}
\end{figure*}

\begin{figure}
    \centering
    \includegraphics[width=\linewidth]{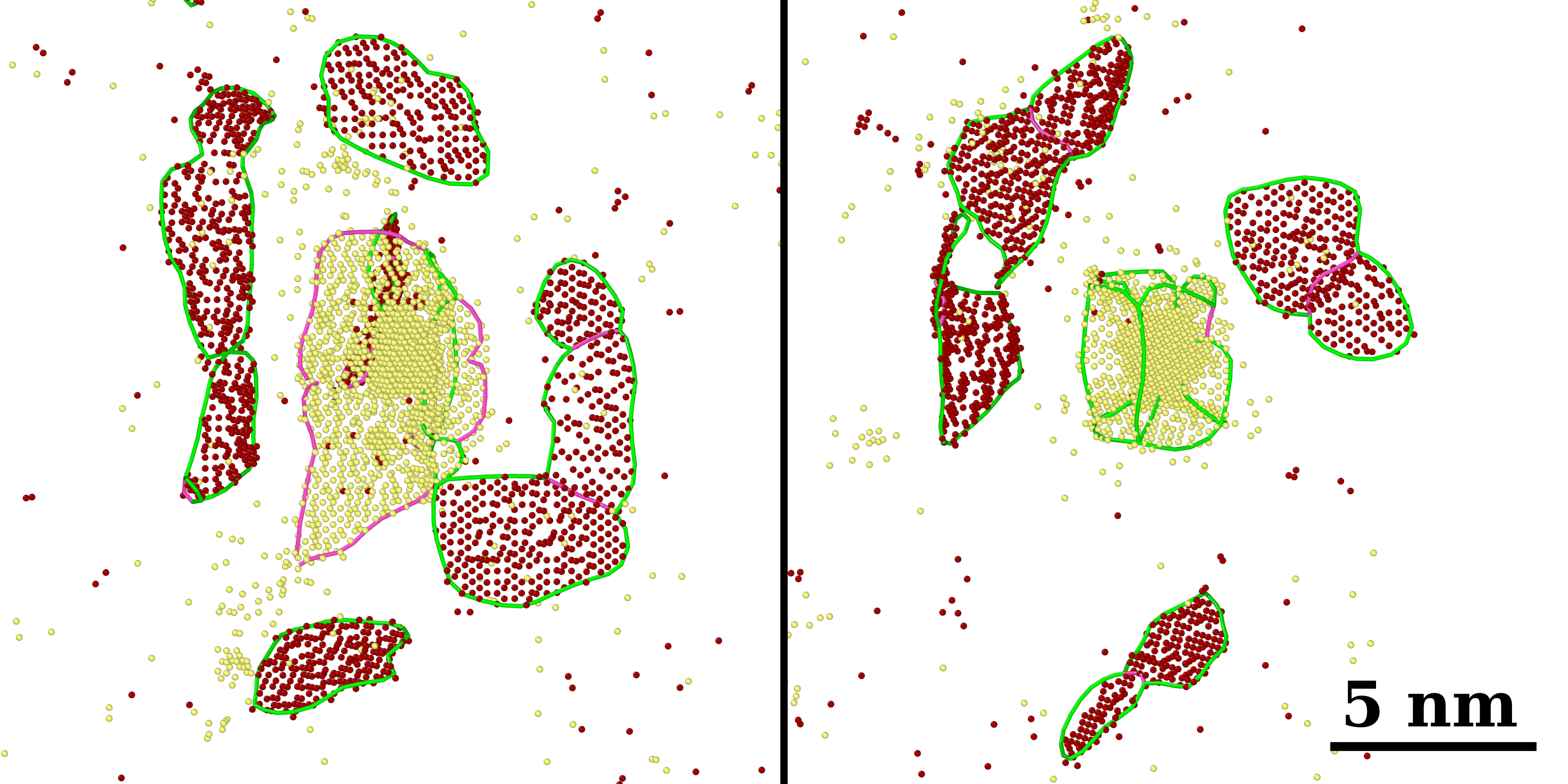}
    \caption{Two examples of large vacancy clusters surrounded by interstitial-type dislocation loops produced by 2 MeV cascades. Vacancies are colored yellow, self-interstitial atoms red, 1/2\hkl<111> dislocation lines green and \hkl<100> dislocations pink.}
    \label{fig:snapshots}
\end{figure}

\begin{figure}
    \centering
    \includegraphics[width=\columnwidth]{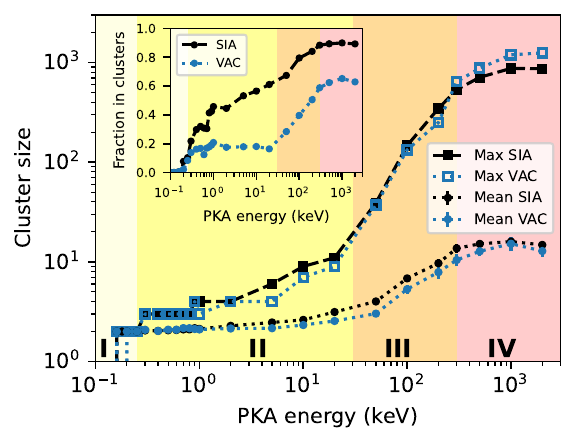}
    \caption{Defect cluster statistics as functions of PKA energy: maximum and mean cluster sizes of vacancy and self-interstitial clusters, and the fraction of vacancies and self-interstitials in clusters (inset figure). The figures are by background color grouped into the same four regimes as discussed in the text.}
    \label{fig:clust}
\end{figure}

Fig.~\ref{fig:casc} shows typical collision cascades spanning three orders of magnitude in recoil energy. By analysing the cascade evolution, morphology, and the final surviving primary damage, we group the damage production into four regimes as a function of energy. The four energy regimes are visualised in the number of Frenkel pairs in Fig.~\ref{fig:nvacs}. The first two regimes (regimes I and II) follow the arc-dpa model~\cite{nordlund_improving_2018} with the low-energy correction~\cite{yang_full_2021} and are well understood. Regime I spans near-threshold displacement energies (TDE). The lowest TDE in W is $42\pm1$ eV for recoils close to the \hkl<100> directions and slightly higher for \hkl<111> directions, measured experimentally and reproduced well by the machine-learned potential used here~\cite{maury_frenkel_1978, byggmastar_threshold_2024}. The average TDE over all crystal directions is 95 eV and the maximum exceeds 250 eV~\cite{byggmastar_threshold_2024}. For energies around the average TDE, at 80--130 eV, there is a plateau where approximately 0.25 Frenkel pairs are produced on average per recoil, similar to what has been observed experimentally in Cu~\cite{king_threshold_1983}. The production of at least one Frenkel pair on average occurs for energies above 500 eV, highlighting the efficient defect recombination for energies far above TDEs.

The sublinear regime II is characterised by conventional collision cascades with increasingly strong heat spikes but no clear subcascade splitting, exemplified by the 20 keV cascade in Fig.~\ref{fig:casc}. Our data for this regime agrees well with the arc-dpa equation fitted to MD data in Ref.~\cite{nordlund_improving_2018} after increasing the average TDE to 95 eV. 

The superlinear regime III agrees with the data in Refs.~\cite{setyawan_cascade_2015} and \cite{nordlund_improving_2018} and starts at around 20--30 keV. At these energies and above, subcascade splitting starts occurring but most cascades still do not split (see simulated probabilities in inset of Fig.~\ref{fig:nvacs}) or split into subcascades where the heat spikes overlap. We found that cascades that do not split but contain an increasing amount of energy within very dense heat spikes (or closely overlapping heat spikes) produce anomalously large defect clusters and total number of defects compared to cascades the split into subcascades. These dense cascades are frequent enough that they significantly increase the mean number of defects and produce a superlinear trend instead of the expected (sub)linear trend assumed by the arc-dpa model. The defect production of dense heat spikes is further simulated and analysed from artificial deposited heat spike simulations in the Supplemental document~\cite{SUPP}. Our analysis suggests that there exists a threshold heat spike energy density above which the effect of single heat spikes transition from causing a sublinear trend to a superlinear trend. That is, heat spikes change from promoting recombination of defects from ballistically displaced atoms to the creation of anomalously large defect clusters that are self-immobilised and cannot recombine easily. We found that this transition occurs at 20--30 keV in W (heat spike radii around 3 nm) and marks the start of the superlinear regime III (see the supplemental Fig. S4~\cite{SUPP}).

Examples of anomalously large defect clusters from dense heat spikes are shown in Fig.~\ref{fig:snapshots}. The largest vacancy clusters often consist of a void-like core with dislocation lines attached or individual large dislocation loops. Vacancy dislocation loops with both \hkl<100> and 1/2\hkl<111> Burgers vectors are frequently formed. Self-interstitial clusters are almost exclusively dislocation loops with the 1/2\hkl<111> Burgers vector pushed out from the heat spike core, in line with previous studies~\cite{sand_high-energy_2013,granberg_molecular_2021}. The statistics of the cascade-produced dislocations are shown in the Supplemental document~\cite{SUPP}.

The linear regime IV is for the first time revealed and quantified here. For energies above 300 keV, most cascades split into subcascades that are far enough apart not to overlap. Hence, the effect of the dense regime-III cascades that cause the superlinear trend reaches a maximum corresponding to the maximum subcascade energy. Instead of containing more energy within a dense heat spike, which would create even larger clusters and continue the superlinear trend, the cascades always split into increasingly separated and independent subcascades. As the subcascade splitting probability approaches one (inset of Fig.~\ref{fig:nvacs}), the total number of defects becomes equivalent to a sum of lower-energy cascades and the trend smoothly transitions to the linear trend that is assumed in all damage models and will continue beyond the range simulated here. Further discussion and numerical proof for the transition from a superlinear to a linear regime due to subcascade splitting are provided in the Supplemental document using a recursive Monte Carlo damage model~\cite{SUPP}.

An analytical defect production model that follows the observed trend across all four regimes can be obtained by multiplying the arc-dpa model with an enhancement factor $[1 + \zeta(E_\mathrm{dam})]$ so that the superlinear trend in regime III is reproduced. The enhancement function $\zeta (E_\mathrm{dam})$ should have the shape of a sigmoid function that increases from 0 at the start of regime III and smoothly saturates towards a constant when transitioning towards regime IV, where the subcascade splitting probability approaches one. This constant corresponds to the relative increase in the number of defects compared to the arc-dpa model. We find that the following full-range damage model for the number of defects as a function of damage energy $N_\mathrm{d}(E_\mathrm{dam})$ describes the data well:
\begin{equation}
    N_\mathrm{d}(E_\mathrm{dam}) = 
\begin{cases}
  0 \text{, if $E_\mathrm{dam} < E_\mathrm{TDE}^\mathrm{min}$},\\
  \cfrac{0.8E_\mathrm{dam.}}{2E_\mathrm{TDE}^\mathrm{avg}}  \text{, if $E_\mathrm{TDE}^\mathrm{min}\le E_\mathrm{dam} \le \frac{2E_\mathrm{TDE}^\mathrm{avg}}{0.8}$},\\
  \cfrac{0.8E_\mathrm{dam}}{2E_\mathrm{TDE}^\mathrm{avg}} \xi(E_\mathrm{dam}) \left[ 1 + \zeta(E_\mathrm{dam}) \right] \text{, else.}
\end{cases}
\end{equation}
This combines the arc-dpa equation, the low-energy correction from Ref.~\cite{yang_full_2021}, and the new enhancement function $\zeta$. $E_\mathrm{TDE}^\mathrm{min}$ and $E_\mathrm{TDE}^\mathrm{avg}$ are the minimum and average threshold displacement energies (we use 42 eV and 95 eV for W~\cite{maury_frenkel_1978,byggmastar_threshold_2024}), $\xi$ is the arc-dpa efficiency function~\cite{nordlund_improving_2018}. For $\zeta$ we found that the generalised logistic function reproduces the data well: 
\begin{equation}
  \zeta(E_\mathrm{dam}) =  \cfrac{\alpha}{1 + \exp\left(-\beta{\cfrac{(E_\mathrm{dam} - E_\mathrm{III})}{\sqrt{E_\mathrm{dam} E_\mathrm{III}}}}\right)}.
\end{equation}
$\zeta$ depends on three material-specific parameters $E_\mathrm{III}$, $\alpha$, and $\beta$. The MD data for W is reproduced well with $E_\mathrm{III} = 140$ keV and $\alpha=\beta=2$. $E_\mathrm{III}$ defines the energy at which $\zeta$ is half of its maximum value ($\alpha$) and is hence approximately the mid-point damage energy of the superlinear regime III. $\alpha$ is the maximum value as $E_\mathrm{dam} \rightarrow \infty$ and defines how much higher the regime-IV linear trend is compared to the arc-dpa model. $\beta$ controls the width of the transition. The arc-dpa efficiency function contains two material-specific parameters, which together with the three new parameters must be estimated or fitted when applying the model to other materials.

The four regimes become more clear when analysing the statistics of defect clusters produced by cascades at different energies, summarised in Fig.~\ref{fig:clust}. The formation of larger and larger vacancy and self-interstitial clusters in the superlinear regime III is clear from Fig.~\ref{fig:clust}. When reaching regime IV, the distribution of cluster sizes, including the mean and the maximum sizes, plateaus since the cascades split into smaller fully separated subcascades equivalent to lower-energy cascades. The fraction of vacancies and interstitials in clusters, shown in the inset of Fig.~\ref{fig:clust}, shows a similar overall trend, with a plateau or slowly increasing clustered fraction at low energies, a rapid increase in regime III and plateauing to a constant in regime IV.

Fig.~\ref{fig:clust} suggests that the largest clusters that frequently can form directly from collision cascades in pristine bulk tungsten is around 1000 vacancies or self-interstitial atoms. Clusters of this size are 4--10 nm in width depending on morphology. Previous simulations and experiments have confirmed that the frequency of forming clusters of given sizes follow power laws for small cluster sizes~\cite{sand_high-energy_2013,yi_direct_2015}. Sand et al. later developed and experimentally validated a correction to the power law for large clusters that at the time were not reachable by simulations~\cite{sand_cascade_2017}. That model predicts an upper limit to the cluster size, which our simulations now directly confirm. The upper limit is also supported by experiments by Ciupiński et al., who used 20 MeV self-ion irradiation and observed dislocation loops only around 5 nm in size at low dose~\cite{ciupinski_tem_2013}, translating to similar maximum cluster sizes that we observe here. Cluster distributions from the simulations for all energies are provided in the Supplemental document~\cite{SUPP}. We also note that while the mean size of the clusters is always higher for interstitials, the largest vacancy clusters become larger than the largest interstitial clusters at the highest damage energies. The opposite is true for low energies, and regime III is the transition region also in this case. The clustered fractions in the inset of Fig.~\ref{fig:clust} reflect the same trend. At low energies, around half of interstitials formed directly by the cascades are in clusters while only 15--20\% of vacancies are clustered. The transition regime III marks a sharp increase in both, and at high energies almost all interstitials and around 60\% of vacancies are in clusters, most of them in large clusters similar to those in Fig.~\ref{fig:snapshots}.

To conclude, we have here provided accurate simulated data for the primary radiation damage of tungsten across all relevant energies, from the threshold displacement energy 40 eV up to 2 MeV, with a simple linear trend that extrapolates to even higher energies. Strikingly, the final linear regime IV revealed and characterised here starts at around 300 keV, coinciding with the highest recoil energy from 14.1 MeV fusion neutrons. This can be considered a fortunate coincidence, because it means that even in high-energy heavy-ion irradiation experiments with recoils beyond fusion relevancy, the induced collision cascades split and behave similarly to fusion-relevant recoils and the extent of the primary damage can be linearly extrapolated from the high-energy data of Figs.~\ref{fig:nvacs} and \ref{fig:clust}. Finally we have provided a revised damage model for the number of defects in the primary damage that reproduces the observed trends across all four regimes, from near-threshold energies to sublinear, superlinear, and the final linear trend.

\textit{Acknowledgments.} JB and FG acknowledge funding from the Research council of Finland through the OCRAMLIP (grant number 354234) and the DEVHIS (grant number 340538) projects, respectively. This work has been partially carried out under the EUROfusion E-TASC Advanced Computing Hub project. This work has partially been carried out within the framework of the EUROfusion Consortium, funded by the European Union via the Euratom Research and Training Programme (Grant Agreement No 101052200 — EUROfusion). Views and opinions expressed are however those of the author(s) only and do not necessarily reflect those of the European Union or the European Commission. Neither the European Union nor the European Commission can be held responsible for them. We acknowledge University of Helsinki, Finland for awarding this project access to the LUMI supercomputer, owned by the EuroHPC Joint Undertaking, hosted by CSC (Finland) and the LUMI consortium.

%\section*{Acknowledgements}

%\section*{References}
%\bibliography{mybib}


\end{document}

\title{Supplemental information of: Four regimes of primary radiation damage in tungsten}
%\shorttitle{Test}

\author{J. Byggmästar}
\thanks{Corresponding author}
\email{jesper.byggmastar@helsinki.fi}
\affiliation{Department of Physics, P.O. Box 43, FI-00014 University of Helsinki, Finland}
\author{V-M. Yli-Suutala}
\affiliation{Faculty of Science and Engineering, Åbo Akademi University, Vattenborgsvägen 3, FI-20500 Åbo, Finland}
\affiliation{Department of Physics, P.O. Box 43, FI-00014 University of Helsinki, Finland}
\author{A. Fellman}
\affiliation{Department of Physics, P.O. Box 43, FI-00014 University of Helsinki, Finland}
\author{J. Åström}
\affiliation{CSC -- IT Center for Science Ltd., P.O. Box 405, 02101 Espoo, Finland}
\affiliation{Department of Physics, P.O. Box 43, FI-00014 University of Helsinki, Finland}
\author{J. Westerholm}
\affiliation{Faculty of Science and Engineering, Åbo Akademi University, Vattenborgsvägen 3, FI-20500 Åbo, Finland}
\author{F. Granberg}
\affiliation{Department of Physics, P.O. Box 43, FI-00014 University of Helsinki, Finland}

\date{\today}

\maketitle

\section{GPU implementation of \MakeLowercase{tab}GAP}

We ported the \textsc{lammps} component of tabGAP (\url{https://gitlab.com/jezper/tabgap}) to GPUs using the Kokkos library~\cite{9485033}. The force computation is parallelized over the atoms. For simulations of one element, the Kokkos version of the code is 1.6 times faster when run on one GCD of an AMD MI250x GPU, compared to the original non-GPU code run on a full CPU node on LUMI with 128 AMD EPYC 7763 cores.

The force computation is performed on three atoms at once using a three-dimensional spline interpolation. The elements of the three atoms, along with their relative positions, determine which spline coefficients are to be used to compute the force on them. When more chemical elements are added to a simulation utilizing tabGAP as the interatomic potential, the performance degrades. This is because reading more different spline coefficients from memory leads to worse cache utilization. To help improve the performance of simulations with multiple elements, the neighbor list of each atom is sorted by element. A list of atoms sorted by element is also constructed. The force computation then loops over the sorted atoms and neighbors. Sorting the neighbors and atoms improves performance by 1.4x for simulations of five elements, each with an equal number of atoms distributed randomly in the simulation box. The code is available open source (\url{https://gitlab.com/jezper/tabgap}) and easily compiled with \textsc{lammps}.

We also ported the friction force electronic stopping routine of \textsc{lammps} to Kokkos (\texttt{fix electron/stopping}) to allow all frequent computations needed in cascade simulations to be run on GPU. This code is already merged into the main repository of \textsc{lammps} \url{https://github.com/lammps/lammps}.

\section{Supplemental animations}

The supplemental material contains animations of the representative cascade simulations showed in Fig. 1 in the main text. In the files with \texttt{ekin} in the file names, atoms with high kinetic energies are shown, to visualise the cascade structure and pressure pulses. In the files with \texttt{ws} in the file names, vacancies (blue) and interstitial atoms (yellow) identified by Wigner-Seitz analysis are shown, to visualise the formation of the final defect clusters. Note that due to the adaptive time step used in the simulations, the simulation time interval between animated frames is not constant (short time span in the beginning due to strong collisions and faster time evolution during the heat spike and until the end).

\clearpage
\section{Cascade results}

Below, the key data from the cascade simulations are given in copyable plain text. The columns are: PKA energy in keV \texttt{E\_PKA}, damage energy \texttt{E\_dam}, average number of Frenkel pairs \texttt{N\_FP}, standard deviation of \texttt{N\_FP}, standard error of \texttt{N\_FP}, maximum interstitial cluster size \texttt{max\_sia}$^\mathrm{a}$, mean interstitial cluster size \texttt{mean\_sia}, maximum vacancy cluster size \texttt{max\_vac}, mean vacancy cluster size \texttt{mean\_vac}, fraction of interstitials in clusters \texttt{cf\_sia}, fraction of vacancies in clusters \texttt{cf\_vac}, number of simulated recoils \texttt{N\_sim}, number of atoms per recoil in the system \texttt{N\_atoms}$^\mathrm{b}$, and average computational cost (GPU-hours) per cascade \texttt{GPUh}. The GPUs were AMD MI250x GPUs on the LUMI supercomputer in all simulations.

$^\mathrm{a}$A cluster is defined as containing two or more vacancies or interstitials (a maximum or mean size of 1 means that all defects are isolated).

$^\mathrm{b}$Number of atoms per recoil. For energies 40--250 eV, 1000 recoils were placed on a grid and simulated simultaneously in a box of 54 million atoms. For 300--1000 eV, 125 recoils were simulated simultaneously in a box of 16 million atoms. For all higher energies, each recoil is simulated in a separate system of the given size.

\begin{center}
%\begin{minipage}{\paperwidth}
%\centering
\begin{footnotesize}
\begin{verbatim}
E_PKA (keV)  E_dam (keV)  N_FP     stddev  stderr  max_sia  mean_sia  max_vac  mean_vac  cf_sia cf_vac N_sim N_atoms    GPUh  
0.04         0.04         0.002    0.00    0.00    1        1.000     1        1.000     0.000  0.000  1000  5.400e+04  0.007 
0.05         0.05         0.042    0.00    0.00    1        1.000     1        1.000     0.000  0.000  1000  5.400e+04  0.007 
0.06         0.06         0.137    0.00    0.00    1        1.000     1        1.000     0.000  0.000  1000  5.400e+04  0.007 
0.07         0.07         0.198    0.00    0.00    1        1.000     1        1.000     0.000  0.000  1000  5.400e+04  0.007 
0.08         0.08         0.248    0.00    0.00    1        1.000     1        1.000     0.000  0.000  1000  5.400e+04  0.007 
0.09         0.08         0.253    0.00    0.00    1        1.000     1        1.000     0.000  0.000  1000  5.400e+04  0.007 
0.10         0.09         0.231    0.00    0.00    1        1.000     1        1.000     0.000  0.000  1000  5.400e+04  0.007 
0.12         0.11         0.245    0.00    0.00    1        1.000     1        1.000     0.000  0.000  1000  5.400e+04  0.007 
0.14         0.13         0.277    0.00    0.00    1        1.000     1        1.000     0.000  0.000  1000  5.400e+04  0.007 
0.16         0.15         0.322    0.00    0.00    2        2.000     2        2.000     0.006  0.006  1000  5.400e+04  0.007 
0.18         0.17         0.394    0.00    0.00    2        2.000     1        1.000     0.010  0.000  1000  5.400e+04  0.007 
0.20         0.18         0.464    0.00    0.00    2        2.000     2        2.000     0.078  0.022  1000  5.400e+04  0.007 
0.25         0.23         0.62     0.00    0.00    2        2.000     2        2.000     0.103  0.084  1000  5.400e+04  0.007 
0.30         0.27         0.752    0.00    0.00    3        2.050     3        2.080     0.218  0.138  500   1.280e+05  0.017 
0.40         0.36         0.926    0.00    0.00    3        2.029     3        2.027     0.298  0.162  500   1.280e+05  0.017 
0.50         0.44         1.144    0.00    0.00    3        2.056     3        2.087     0.320  0.168  500   1.280e+05  0.017 
0.60         0.53         1.18     0.00    0.00    3        2.068     3        2.086     0.308  0.124  500   1.280e+05  0.017 
0.70         0.61         1.254    0.00    0.00    3        2.099     3        2.163     0.305  0.169  500   1.280e+05  0.017 
0.80         0.70         1.524    0.00    0.00    3        2.100     3        2.159     0.414  0.178  500   1.280e+05  0.017 
0.90         0.78         1.666    0.00    0.00    4        2.159     4        2.120     0.425  0.191  500   1.280e+05  0.017 
1.00         0.86         1.854    0.00    0.00    4        2.126     3        2.098     0.458  0.208  500   1.280e+05  0.017 
2.00         1.69         2.57     1.17    0.12    4        2.280     4        2.143     0.445  0.175  100   1.280e+05  0.020 
5.00         4.12         4.59     1.97    0.20    6        2.465     4        2.158     0.533  0.179  100   1.280e+05  0.021 
10.00        8.06         7.59     2.93    0.29    9        2.616     7        2.322     0.565  0.181  100   1.024e+06  0.142 
20.00        15.74        12.43    4.38    0.44    11       3.136     9        2.550     0.611  0.164  100   1.024e+06  0.146 
50.00        38.10        29.5     8.00    0.80    39       4.004     37       3.022     0.674  0.284  100   8.192e+06  1.250 
100.00       74.54        71.54    28.20   2.82    147      6.824     133      5.318     0.795  0.397  100   1.600e+07  2.750 
200.00       144.18       173.04   75.45   10.67   346      9.684     250      7.879     0.841  0.505  50    5.400e+07  10.667
300.00       211.94       309.7    154.72  21.88   534      13.671    642      10.438    0.884  0.588  50    1.280e+08  32.000
500.00       347.41       552.76   216.74  30.65   711      15.157    877      12.754    0.894  0.624  50    1.280e+08  40.000
1000.00      655.98       1145.2   433.27  96.88   874      15.978    1195     15.200    0.900  0.652  20    1.024e+09  725.00
2000.00      1198.25      1967.45  588.34  131.56  867      14.799    1251     12.839    0.893  0.629  20    1.024e+09  1280.0
\end{verbatim}
\end{footnotesize}
%\end{minipage}
\end{center}

\clearpage
\section{Dislocations}

\begin{figure}[h]
    \centering
    \includegraphics[width=0.8\columnwidth]{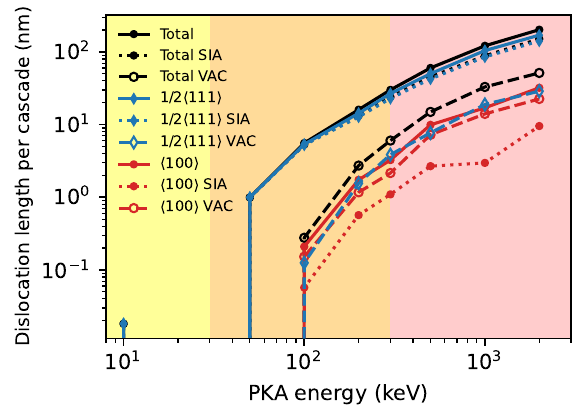}
    \caption{Average total dislocation lengths produced in cascades as function of PKA energy. The background colors divide the energy range into the different regimes described in the main text (the lowest regime produces no dislocations and is not visible here).}
    \label{fig:disl}
\end{figure}

Fig.~\ref{fig:disl} shows the total dislocation length per cascade as a function of PKA energy. The lengths are further divided into vacancy and interstitial-type dislocations, and the two Burgers vectors. The dislocations are identified with the DXA algorithm~\cite{stukowski_automated_2012}.

\section{Cluster distributions}

\begin{figure*}
    \centering
    \includegraphics[width=\linewidth]{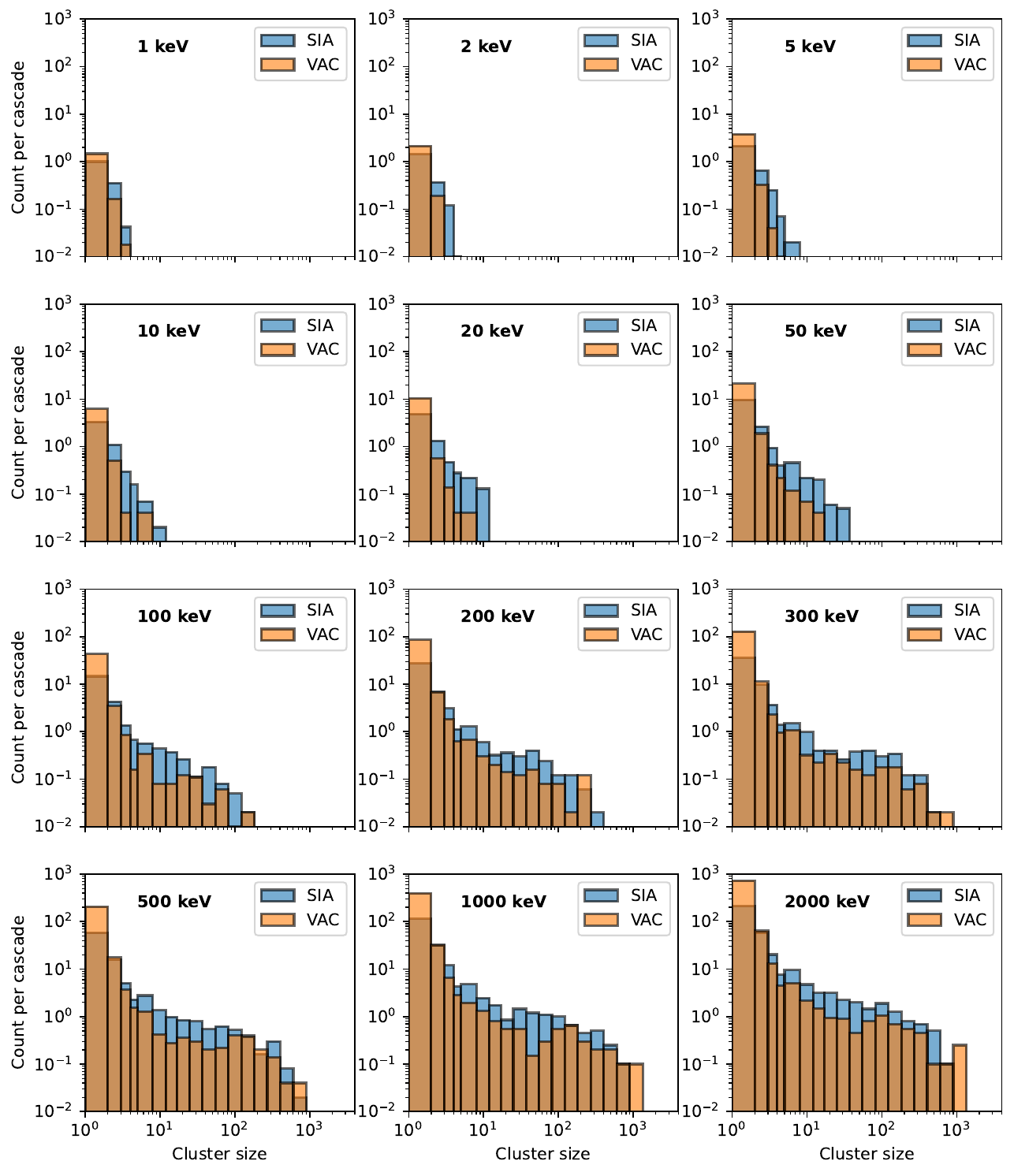}
    \caption{Histograms of cluster sizes for all PKA energies 1 keV and above for both vacancy and self-interstitial atom clusters.}
    \label{fig:clust_distr}
\end{figure*}

Fig.~\ref{fig:clust_distr} shows histograms of cluster sizes for vacancy and interstitial clusters for all simulated PKA energies above and including 1 keV.

\clearpage
\section{Artificial heat spike simulations}

As explained in the Letter, the transition to a superlinear trend in regime III is according to our simulations the extreme defect production of compact high-energy cascades. Such dense cascades are possible because the subcascade splitting threshold in W is so high that cascades from PKA energies of hundreds of keV have significant probabilities to \textit{not} split into subcascades or split into subcascades that are close enough that their heat spikes overlap. Some high-energy cascades that do not split into subcascades remain very compact, leading to an extreme amount of energy concentrated in a local region. Such dense compact cascades produce far more defects than cascades that split into subcascades (including when subcascades overlap). The high number of defects is due to the formation of anomalously large defect clusters in forms of central vacancy clusters surrounded by large interstitial loops, like those illustrated in Fig. 3 in the manuscript. These clusters are often self-immobilized, preventing recombination. It is the defect production of these high-energy dense cascades that drive the defect production trend to the superlinear regime III.

To verify that the above observation is true, we performed simulations of artificial heat spikes where we deposit a given amount of kinetic energy (damage energy) in a spherical region. Such heat spikes represent and at least qualitatively correspond to an ideal limiting case where all kinetic energy from the initial PKA is deposited in a single spherical region. The simulations were done by first relaxing a single-crystal W lattice in the $NPT$ ensemble at 300 K and zero pressure for 8 ps. After that, kinetic energy corresponding to a desired cascade damage energy was deposited to a central spherical region of atoms in the lattice. The kinetic energy was deposited as a temperature using the \texttt{velocity create} command in \textsc{lammps}. The deposition temperature was converted from total kinetic energy as $E = \frac{3}{2} N k_\mathrm{B}T$ (and multiplying by two to account for the equipartition of kinetic and potential energy). The number of atoms in the hot sphere is $N$, which defines the size of the heat spike for a given damage energy. $N$ was calculated by using the effective energy for melting in cascades, estimated to 2.7 eV/atom for W in Ref.~\cite{boleininger_microstructure_2023}. In practice in the simulation, we defined a spherical region with a radius that corresponds approximately to the calculated number of ``liquid'' atoms. After the energy deposition, the simulation is run in the same way as cascades for 60 ps in $NVE$ with a border thermostat and adaptive time step. Electronic stopping was not needed since the deposited energies are small.

To obtain the defect production trend, we simulated 20 independent simulations (with different initial seeds) for damage energies 2, 10, 20, 50, 100, and 200 keV. The mean number of defects were obtained from Wigner-Seitz analysis, in the same manner as for the cascade simulations.

\begin{figure}[h]
    \centering
    \includegraphics[width=0.15\linewidth]{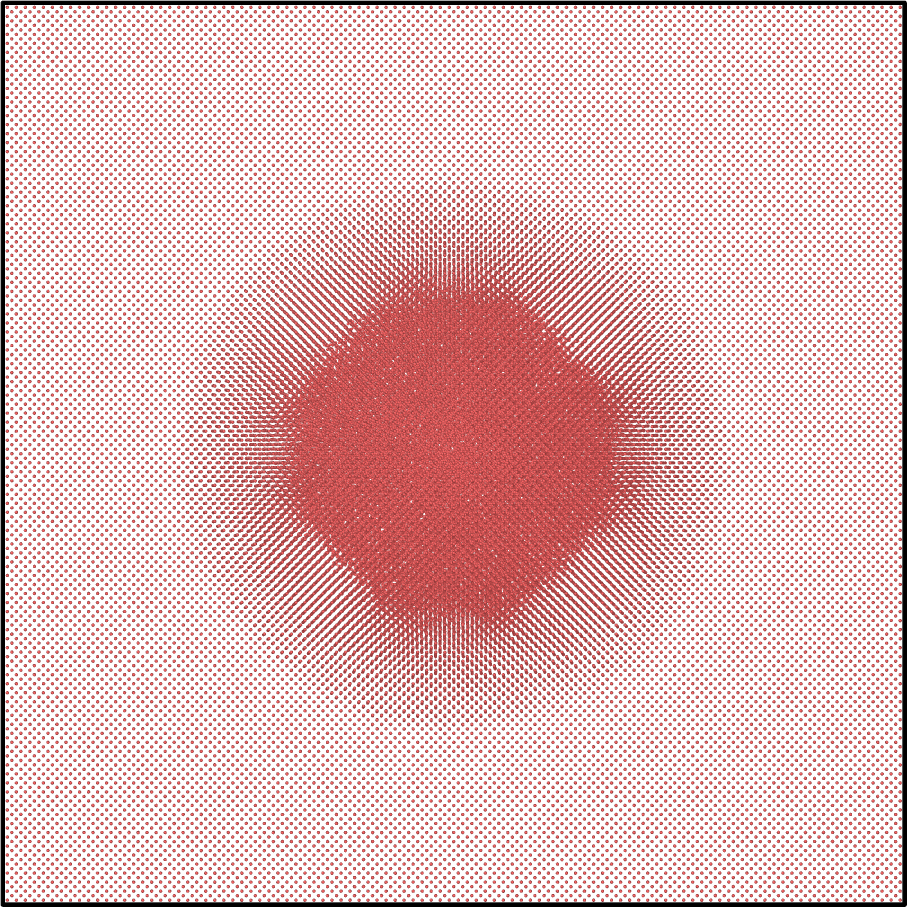}
    \includegraphics[width=0.15\linewidth]{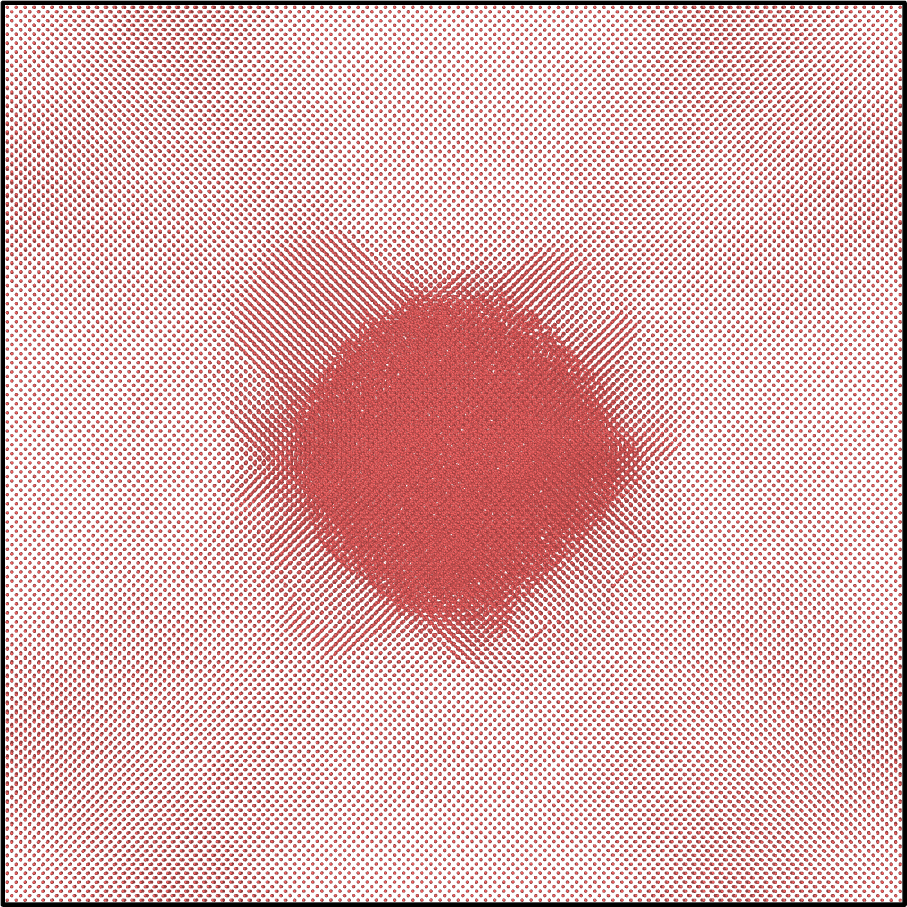}
    \includegraphics[width=0.15\linewidth]{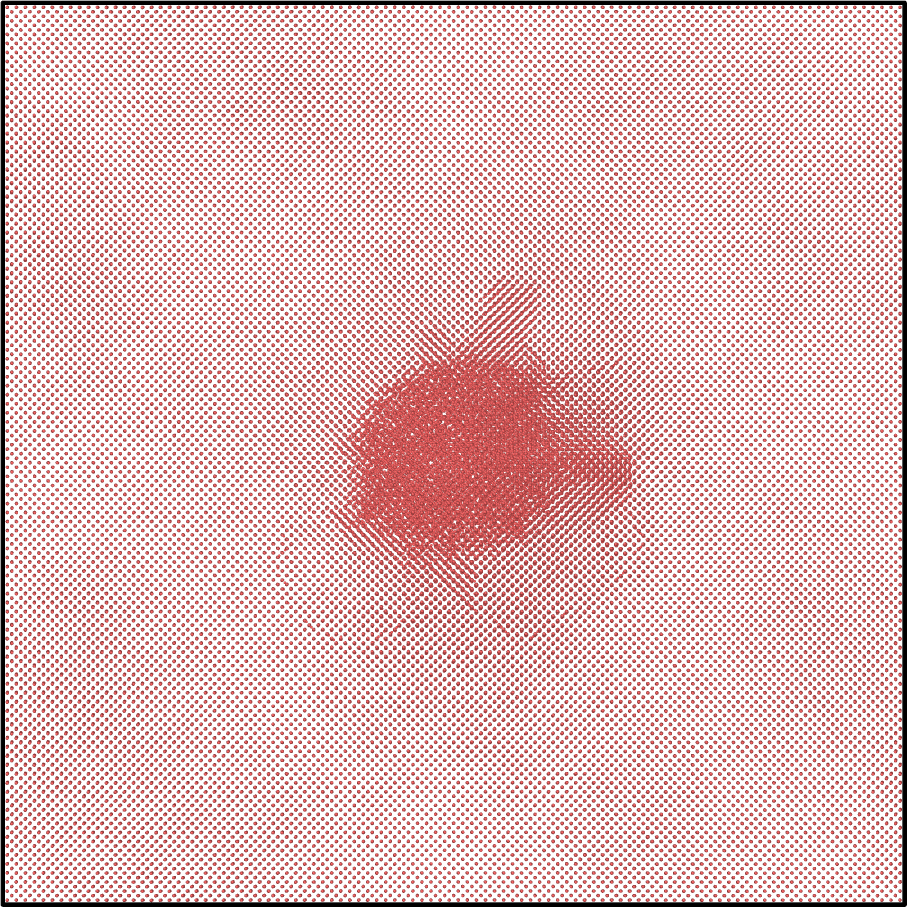}
    \includegraphics[width=0.15\linewidth]{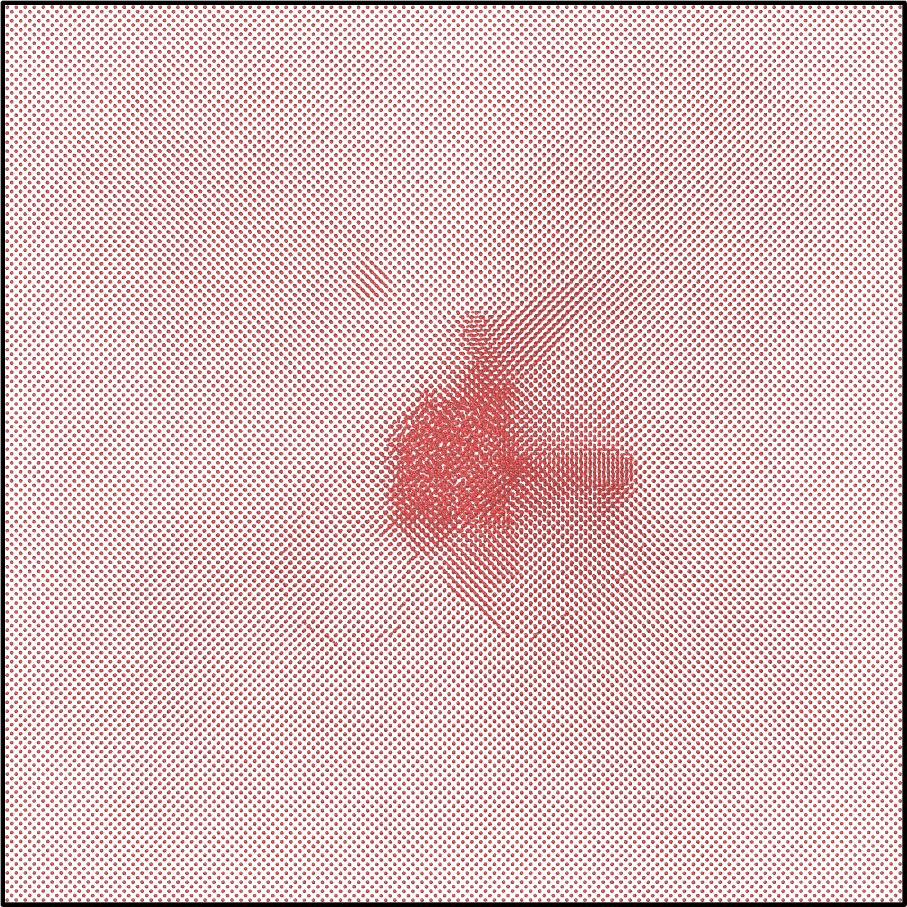}
    \includegraphics[width=0.15\linewidth]{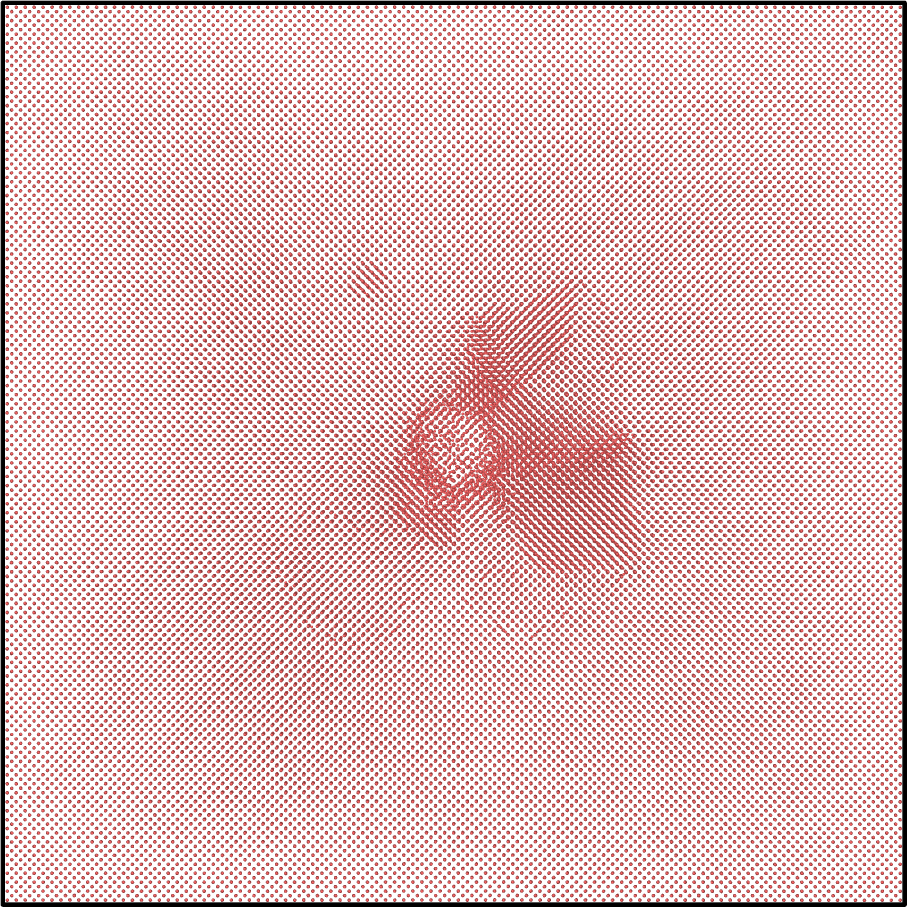}
    \includegraphics[width=0.15\linewidth]{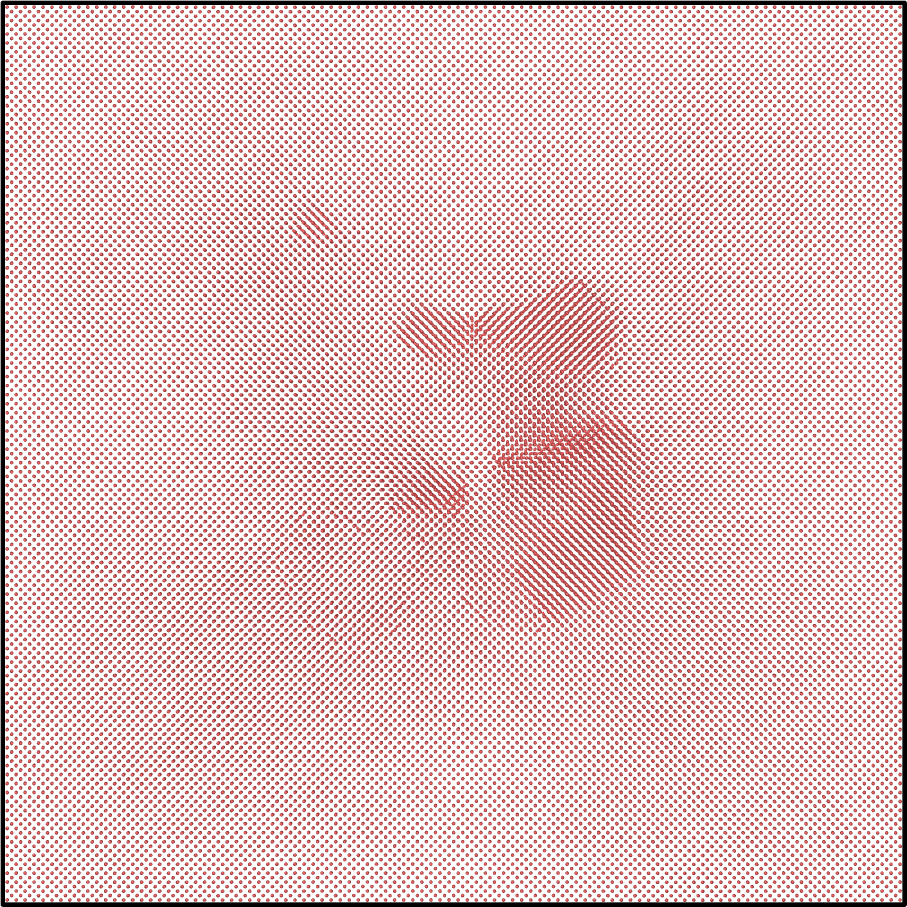}
    \includegraphics[width=0.5\linewidth]{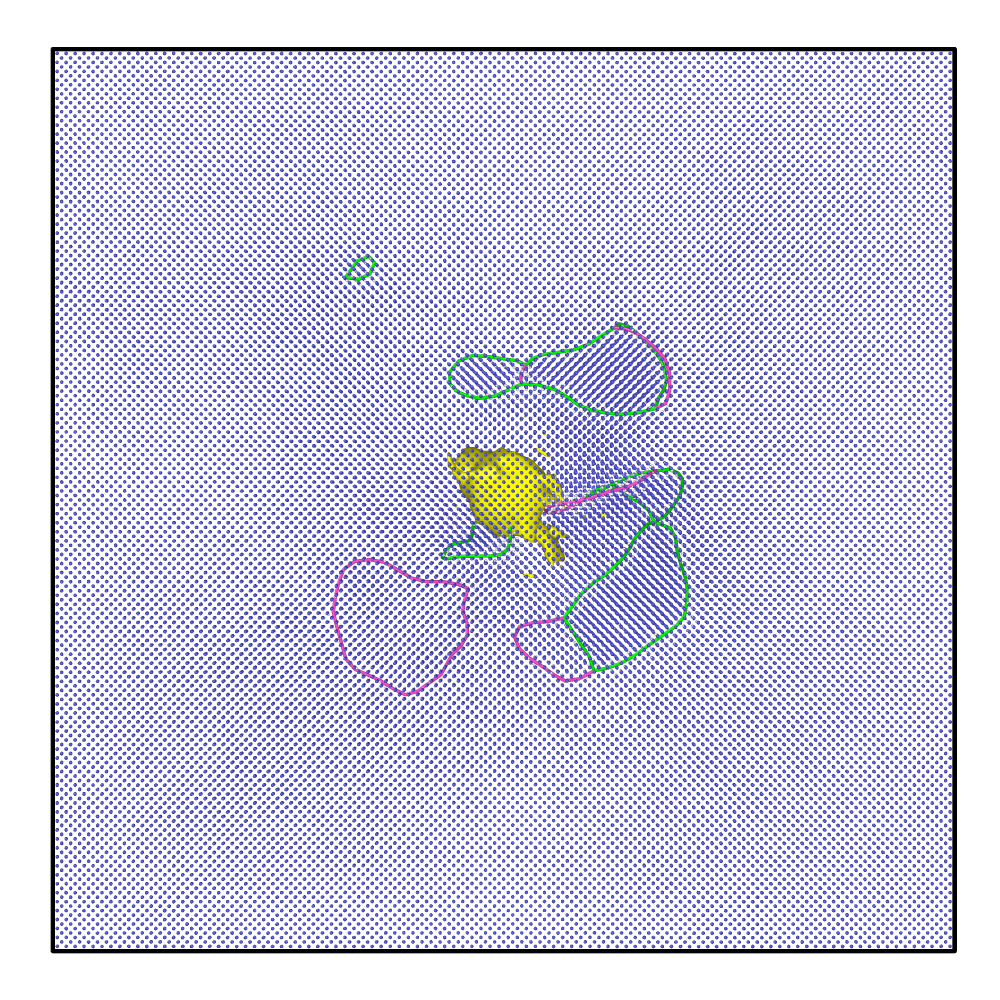}
    \caption{Example of a 100 keV artificial heat spike simulation. The top row shows a time series of snapshots from the initial deposition until complete recrystallisation with residual defects. The bottom snapshots shows the final defect structure with a central void (yellow) surrounded by interstitial dislocation loops.}
    \label{fig:ball_snaps}
\end{figure}

\begin{figure}[h]
    \centering
    \includegraphics[width=0.6\linewidth]{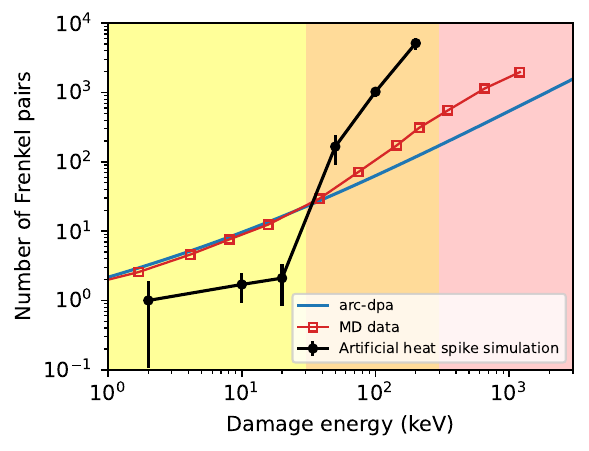}
    \caption{Defect production of artificially deposited spherical heat spikes compared to arc-dpa and the MD cascade data.}
    \label{fig:Rball}
\end{figure}

Snapshots from a representative simulation at 100 keV damage energy are shown in Fig.~\ref{fig:ball_snaps}. The defect production as a function of the deposited damage energy is shown in Fig.~\ref{fig:Rball}. The trend is compared to the arc-dpa model and the MD cascade data. The results show that at very low energies, the artificial heat spikes show very efficient recombination and create much fewer defects than the arc-dpa model and MD cascade data shows. However, at energies above 30 keV, the artificial heat spikes show qualitatively the same phenomenon as explained for dense high-energy cascades above: larger and larger vacancy clusters surrounded by interstitial loops of corresponding size are created in a way that produces a superlinear defect production trend. Remarkably, the critical energy where the defect production from spherical heat spikes exceed the expected arc-dpa trend is at 30 keV, coinciding with the transition from near-linear regime II to superlinear regime III in the MD cascade data. These results strengthen the conclusion that it is the occurrence of dense energetic cascades that is responsible for the superlinear trend. The artificial heat spike data in Fig.~\ref{fig:Rball} always clearly overestimate the MD cascade data in regime III, which is expected since near-spherical cascades are rare and can be assumed to be an upper limit in the defect production from real cascades. If cascades would never split into subcascades, the artificial heat spikes in Fig.~\ref{fig:Rball} would indicate that the defect production trend would remain superlinear. Nevertheless, as we demonstrate in the next section, the increasing probability for subcascade splitting drives the trend back to linear.

\clearpage
\section{Recursive Monte Carlo damage model}

\begin{figure}[h]
    \centering
    \includegraphics[width=0.6\linewidth]{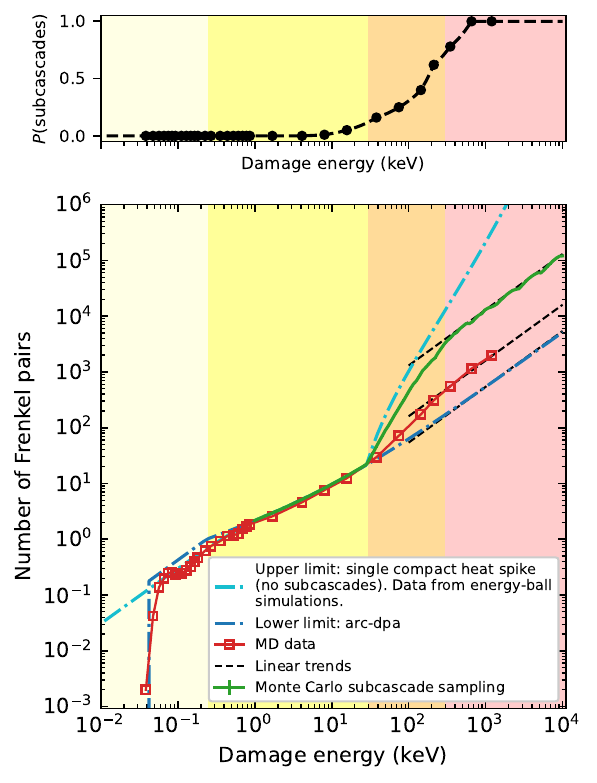}
    \caption{Defect production predicted by the Monte Carlo sampling compared to arc-dpa, MD cascade data, and a spline fit to the artificially deposited heat spike data from Fig.~\ref{fig:Rball}.}
    \label{fig:MC}
\end{figure}

All common damage models (like NRT-dpa and arc-dpa) are constructed so that when the cascades split into subcascades, the defect production is linear because the subcascades can be seen as independent cascades of lower energy. This is also what we observe in our MD simulations. To demonstrate this and reproduce the observed MD trend, we designed and implemented a simple recursive Monte Carlo defect production algorithm that mimics the observed behaviour of cascades. The algorithm relies on the cumulative probability distribution of subcascade splitting $P_\mathrm{sub}(E)$, which we directly obtained from the cascade simulations. $P_\mathrm{sub}(E)$ is shown in Fig.~\ref{fig:MC} (top) and also as an inset in Fig. 2 of the manuscript.

The Monte Carlo algorithm is:
\begin{itemize}
    \item For a given damage energy $E$:
    \begin{enumerate}
        \item Decide by a Monte Carlo step if cascade splits into two subcascades by sampling $P_\mathrm{sub}(E)$ (i.e. draw random number $u \in [0, 1)$ and compare to $P_\mathrm{sub}(E)$).
        \item If $u < P_\mathrm{sub}(E)$, cascade splits to two subcascades:
        \begin{itemize}
            \item Split damage energy $E \rightarrow E_1 + E_2$. [See Note A].
            \item Recursively Monte-Carlo-split $E_1$ and $E_2$ into further subcascades by recursively calling steps 1-2 (i.e. both $E_1$ and $E_2$ may split into subcascades, which in turn may split further, etc.).
        \end{itemize}
        \item Collect the damage energies $E_i$ of all subcascades.
        \item Sample number of defects $N_i$ produced by each subcascade of energy $E_i$:
        \begin{itemize}
            \item if $E_i < 30$ keV: $N_i \leftarrow$ arc-dpa model
            \item else: random number between arc-dpa and artificial heat spike trend. [See Note B].
        \end{itemize}
        \item Total number of defects $N_\mathrm{def} = \sum_i N_i$.
        \item Repeat $M$ times for a given $E$ and return average $N_\mathrm{def}$. In our case $M=1000$ was used for the results.
    \end{enumerate}
\end{itemize}

\begin{itemize}
    \item Note A: In the results shown, we simply always split the energy in half $E_1 = E_2 = E/2$, but we tested different ways of splitting and the results are qualitatively the same.
    \item Note B: We also tested different ways of sampling: directly sampling the artificial heat spike trend or sampling a normal distribution instead of uniformly between arc-dpa and the artificial heat spike trend. This only shifts the final linear trend but the transitions to superlinear and linear regimes remain the same. Correctly sampling this regime would reproduce the MD data, but our purpose was only to demonstrate that the trend will become linear at high energies (due to high probability of subcascade splitting).
\end{itemize}

This simple algorithm mimics the defect production from cascades. At energies below 30 keV, for which according to MD cascades never split into subcascades, the algorithm directly and exactly samples the arc-dpa function. At higher energies, the probability for subcascade splitting increases. Subcascades with energies above 30 keV may produce any number of defects between arc-dpa and the extremely superlinear trend from the artificial deposited heat spike simulations described previously. This produces the superlinear trend that we see from MD cascades, although it is exaggerated in this simplistic algorithm. At energies above 500 keV, the probability of subcascade splitting approaches 1, meaning that the superlinear effect reaches its maximum. For MeV-range energies, the recursively created subcascades from the Monte Carlo algorithm still sample the 30--500 keV superlinear trend, but the trend for the total number of defects becomes linear. Hence, the observed MD cascade trend is reproduced and it proves that even with a superlinear regime, the final high-energy trend will become linear when cascades split into subcascades. 

\begin{figure}[h]
    \centering
    \includegraphics[width=0.5\linewidth]{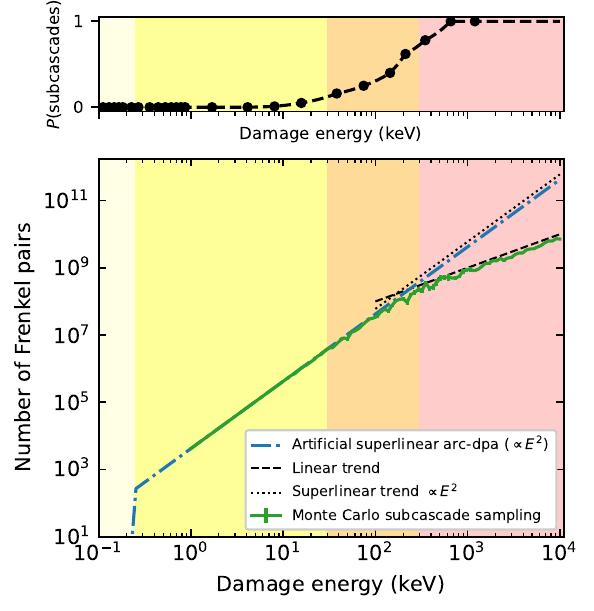}
    \caption{Defect production predicted by the Monte Carlo subcascade sampling from a superlinear damage model, proving that the trend becomes linear when the probability of subcascade splitting approaches 100\%.}
    \label{fig:MCsuper}
\end{figure}

\begin{figure}[h]
    \centering
    \includegraphics[width=0.7\linewidth]{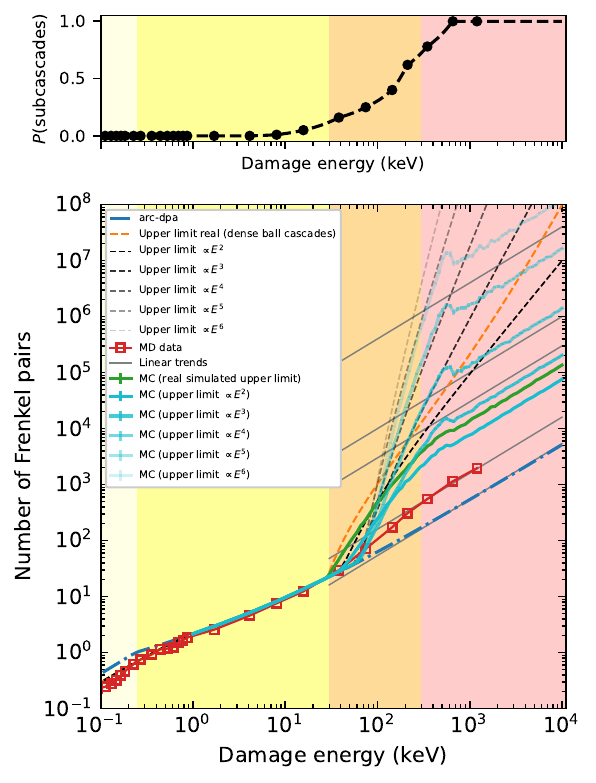}
    \caption{Defect production predicted by the Monte Carlo (MC) subcascade sampling with different artificial functional forms (polynomials) for the superlinear upper limit in defect production, proving that the trend becomes linear when the probability of subcascade splitting approaches 100\%.}
    \label{fig:upperMC}
\end{figure}

The last point is true even if the entire energy range is superlinear, i.e. if we replace arc-dpa with a superlinear function. It is also true regardless of the functional form of the superlinear regime. We confirmed both these cases with the Monte Carlo code with (1) an artificial arc-dpa model where the energy dependence is changed from linear to $\propto E^2$ and (2) by replacing the simulated heat spike upper limit in the superlinear regime by artificial increasingly steep polynomial functions. The results are shown in Figs.~\ref{fig:MCsuper} and \ref{fig:upperMC}, demonstrating transitions to linear trends once the subcascade splitting probability approaches 1 in all cases.

\clearpage
\section{Logistic enhancement function $\zeta$}

\begin{figure}[h]
    \centering
    \includegraphics[width=0.49\linewidth]{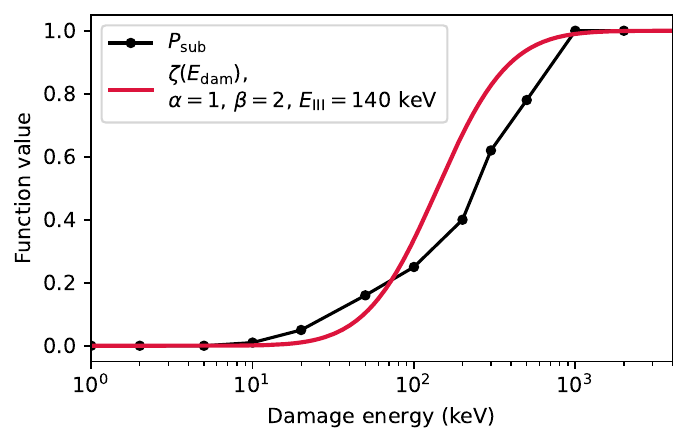}
    \includegraphics[width=0.49\linewidth]{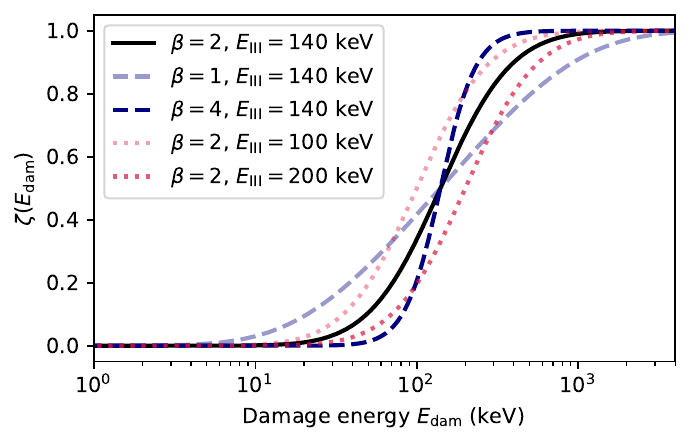}
    \caption{Left: Subcascade probability and the enhancement function $\zeta (E_\mathrm{dam})$ from the new damage model (with the maximum value $\alpha = 1$ for clearer comparison). Right: effect of the material-specific parameters $\beta$ and $E_\mathrm{III}$ on the shape of $\zeta$.}
    \label{fig:zeta}
\end{figure}

Our modified damage model introduces three new material-specific parameters in the enhancement function $\zeta$ as discussed in the manuscript. The enhancement function $\zeta (E_\mathrm{dam})$ is related but not necessarily equal to the probability of subcascade splitting, $P_\mathrm{sub} (E_\mathrm{dam})$, shown previously. The $\zeta$ and $P_\mathrm{sub}$ functions are plotted and compared in Fig.~\ref{fig:zeta}. The effects of the parameters on the shape of $\zeta$ are also illustrated in Fig.~\ref{fig:zeta}.

\begin{equation}
  \zeta(E_\mathrm{dam}) =  \cfrac{\alpha}{1 + \exp\left(-\beta{\cfrac{(E_\mathrm{dam} - E_\mathrm{III})}{\sqrt{E_\mathrm{dam} E_\mathrm{III}}}}\right)}.
\end{equation}

\bibliography{mybib}